\documentclass[12pt]{article}

\usepackage{graphicx}
\newcommand{\EQ}{\begin{equation}}
\newcommand{\EN}{\end{equation}}
\newcommand{\bea}{\begin{eqnarray}}
\newcommand{\ena}{\end{eqnarray}}
\newcommand{\vs}[1]{\vspace{#1 mm}}
\newcommand{\hs}[1]{\hspace{#1 mm}}
\renewcommand{\a}{\alpha}

\renewcommand{\d}{\delta}
\newcommand{\e}{\epsilon}

\def\bbox{{\,\lower0.9pt\vbox{\hrule \hbox{\vrule height 0.2 cm
\hskip 0.2 cm \vrule height 0.2 cm}\hrule}\,}}
\newcommand{\dsl}{\pa \kern-0.5em /}

\newcommand{\pa}{\partial}
\renewcommand{\t}{\theta}

\newcommand{\nn}{\nonumber\\}
\newcommand{\p}[1]{(\ref{#1})}
\newcommand{\lan}{\langle}
\newcommand{\ran}{\rangle}

\begin{document}

\begin{titlepage}
\null
\begin{flushright}
OU-HET 401 \\
UT-977 \\
hep-th/0112050
\end{flushright}

\vs{10}
\begin{center}
{\Large\bf Moduli Space and Scattering of D0-Branes in Noncommutative
Super Yang-Mills Theory}
\vs{15}

{\large Masashi Hamanaka\footnote{e-mail address:
    hamanaka@hep-th.phys.s.u-tokyo.ac.jp},
    Yasuyuki Imaizumi\footnote{e-mail address:
    imaizumi@het.phys.sci.osaka-u.ac.jp} and
    Nobuyoshi Ohta\footnote{e-mail address: ohta@phys.sci.osaka-u.ac.jp}}
\vs{15}

${}^1$
{\em Department of Physics, University of Tokyo, Tokyo 113-0033, Japan}

\vs{6}

${}^{2,3}${\em Department of Physics, Osaka University,
Toyonaka, Osaka 560-0043, Japan}

\vs{12}
{\bf Abstract}
\end{center}

We study the moduli space of the D0-brane system on D$p$-branes
realized in the noncommutative super Yang-Mills theory. By examining
the fluctuations around the solitonic solutions generated by solution
generating technique, we confirm the interpretation of the moduli
as the positions of D0-branes on D$p$-branes. Low-energy scattering process
is also examined for two D0-branes. We find that the D0-branes scatter
at right angle for head-on collision in the D0-D4 system. For
D0-D6 and D0-D8 systems we find special solutions which reduce to the
D0-D4 case, giving the same behavior. This suggests that the scattering
at right angle for head-on collision is a universal behavior of this kind
of soliton scatterings.

\end{titlepage}

\clearpage

%%%%%%%%%%%%%%%%%%%%%%%%%%%%%%%%%%%%%%%%%%%%%%%%%%%%%%%%%%%%%%%%%%%%%%%%%%%

\section{Introduction}

D-branes are important solitons in string theory and have revealed not only
various dualities in string theory but also nonperturbative aspects of field
theories~\cite{GK}. Especially D-brane effective theories with background
NS-NS $B$-field have proved to be noncommutative gauge
theories~\cite{CDS,DH,SW}, and this realization has been used to study the
nonperturbative dynamics of noncommutative field theories~\cite{OT}.

Conversely D$p$-branes on D$p'$-branes ($p<p'$) can be described as solitons
in noncommutative theories~\cite{GMS,HKLM,P,Bak,AGMS,HT,Hashimoto}. This allows
investigation of the D-brane dynamics, e.g. tachyon condensation, in terms of
noncommutative gauge theories. By T-duality, these D-brane systems can be
mapped to D0-D$p$ ($p=0,2,4,6,8$) systems in type IIA theory, on which we will
focus in this paper. It is interesting that some non-BPS D-brane systems can
be BPS in appropriate background $B$-field~\cite{CIMM,MPT,W,OK,OTown,FIO}.

Exact noncommutative solitons are very useful to study the dynamics of
D$0$-branes. There are mainly two powerful methods to construct exact
(BPS) solitons in noncommutative gauge theories; ``solution generating
technique''~\cite{HKL} and ADHM construction~\cite{ADHM}.
Solution generating technique is a transformation which keeps
field equations satisfied and generates nontrivial solutions from trivial
ones. ADHM construction is a method based on the one-to-one correspondence
between instanton moduli space and the solution space of ADHM equation.
We can get all instanton solutions by solving ADHM equation.

Noncommutative gauge theories are non-local and have no local observables.
However noncommutative solitons have the moduli parameters which represent
the positions of the solitons. In previous work, evidence is given for
the interpretation of the moduli parameters as the positions of the $k$
solitons in matrix theory~\cite{AGMS}, by use of the Wilson lines~\cite{GN},
by exact Seiberg-Witten map~\cite{HO} and by ADHM construction~\cite{F4,H}.
In this letter, we provide another evidence for this interpretation by
examining the fluctuations around the soliton solutions. The fluctuations
correspond to the open strings between D0-branes. What we find is that the
mass eigenvalues are proportional to the length of the stretched string,
confirming the above interpretation.

The moduli parameters can also be used to study the low-energy D0-brane
scattering on D$p$-branes. This has been discussed for the so-called GMS
solitons~\cite{GMS} in noncommutative scalar field
theories~\cite{LRU,GHS,HLRU,AI}. However, they are approximate solutions
in the leading order in the noncommutativity parameters, and the result is
valid only in the leading approximation in the large noncommutativity
parameters. It is then natural to ask what is the exact result for the soliton
solutions. Here we examine this problem in noncommutative super
Yang-Mills theory, which admits exact BPS soliton solutions. The scattering
is described by geodesic motion, and we obtain the result without
approximation in the noncommutativity parameters.
In particular, we find that the low-energy scattering occurs at right angle
for zero impact parameter, a typical result for soliton scattering including
monopoles~\cite{M,AH,GM}, though the solitons obtained by solution generating
technique scatters trivially. Our results indicate that this feature is
a universal behavior of this kind of soliton scatterings.

\section{Moduli as positions of D0-branes}

We begin by describing the D$0$-D$p$ ($p=2,4,6,8$) systems in type IIA theory
with background constant $B$-field.
The D$p$-brane fills the directions $x_0,\cdots,x_{p}$
and the $B$-field is block-diagonal and is taken to lie
in the directions $(x_1,x_2, \cdots,x_{p-1},x_p)$:
\begin{equation}
B = \mbox{diag}([B_1],\cdots,[B_{p/2}])=
\frac{\e}{2\pi\alpha'}\mbox{diag}([b_1],\cdots,[b_{p/2}]),
\label{bfield}
\end{equation}
where $[B_i]$ and $[b_i]$ $(i=1,\cdots,p/2)$ are $2\times2$ matrices
\begin{equation}
[B_i] =
\left(\begin{array}{cc}
0 & -B_i \\
B_i & 0
\end{array}\right)
=\frac{\e}{2\pi\a'}[b_i]=\frac{\e}{2\pi\a'}
\left(\begin{array}{cc}
0 & -b_i \\
b_i & 0
\end{array}\right).
\label{eq:b-field}
\end{equation}
The metric on the string worldsheet is written as $g_{ab}=\e \d_{ab}\;
(a,b=1,\cdots,p), g_{00}=-1$. Here $\e$ is a parameter to define the zero
slope limit in order to give noncommutative theories~\cite{SW}:
\begin{equation}
\a' \sim \e^{1/2}\to 0, \quad
B:\mbox{ finite}, \quad
b_i \sim \e^{-1/2} \to \infty.
\label{eq:zeroslope}
\end{equation}

In the present letter, we concentrate on the zero slope limit
(\ref{eq:zeroslope}), and consider the corresponding $(p+1)$-dimensional
noncommutative $U(1)$ gauge theory~\cite{CDS,DH,SW}
\begin{eqnarray}
S = -{1\over 4g_{\rm YM}^2 G_s/g_s}
\int\,dt\,d^{p}x\,\sqrt{-G} G^{\mu\lambda} G^{\nu\sigma}
 F_{\mu\nu} * F_{\lambda\sigma},
\label{eq:YM}
\end{eqnarray}
where $g_s$ is the string coupling and satisfies
$g^2_{\rm YM}=(2\pi)^{p-2}(\a')^{(p-3)/2} g_s$ and
\begin{eqnarray}
G_{ab} = g_{ab} - (2\pi\a')^2 (Bg^{-1}B)_{ab}
&\to& \e b^2 \d_{ab}, \nn
G_s = g_s \left(\frac{\mbox{yet}(g+ 2\pi\a' B)}{\mbox{yet}\, g}\right)
^{\frac{1}{2}} &\to& g_s \prod_{i=1}^{p/2} b_i,
\end{eqnarray}
in the zero slope limit. Though we should supplement \p{eq:YM} with
fermionic terms when some supersymmetry is preserved, it is enough to
consider only the bosonic terms for our purpose.

The above representation is in terms of star-product. There is another
formulation of noncommutative theories, known as operator formalism which is
equivalent to the above via Weyl transformation. Let us now switch to
the operator formalism. The noncommutativity of the space coordinates implies
\begin{eqnarray}
[ x^{2i-1},x^{2i} ] =i\t_i,\quad
\theta_i = {2\pi \a' \over \e b_i} = \frac{1}{B_i}, \quad
(i=1,\cdots,p/2),
\end{eqnarray}
where we assume $b_i, \t_i \geq 0$.
We define complex coordinates
\begin{eqnarray}
z_{j}={1\over\sqrt{2}}(x^{2j-1}+ix^{2j}),\quad
\bar z_{j}={1\over\sqrt{2}}(x^{2j-1}-ix^{2j}),
\end{eqnarray}
and creation/annihilation operators $a^\dagger_i =\bar z_i /\sqrt{\t_i}$ and
$a_i = z_i /\sqrt{\t_i}$. In the temporal $A_0 =0$ gauge, we can rewrite
(\ref{eq:YM}) as
\begin{eqnarray}
S &=& -{\prod_{i=1}^{p/2} (2\pi \bar b_i) \over g^2_{\rm NYM}}
\int dt {\cal L} ,\\
{\cal L} &=& {\rm Tr}\left[ \sum_{i=1}^{p/2}
\left(
-\pa_{t}C_i\pa_{t}\bar C_i +\frac{1}{2}\left( \left[ C_i , \bar C_i\right]
+\frac{1}{\bar b_i}\right)^2\right)\right.\nn
&&~~\left. +\sum_{i<j} \left( \left[ C_i , \bar C_j\right]\left[ C_j ,
 \bar C_i \right]+\left[ C_i , C_j\right]\left[ \bar C_j , \bar C_i\right]
\right)\right]
\label{action}
\end{eqnarray}
where we have set $g^2_{\rm NYM} = g^2_{\rm YM}\prod_{i=1}^{p/2} b_i$,
$\bar b_i=\e b_i^2 \t_i = 2\pi\a' b_i$ and
\begin{eqnarray}
C_j = C_{z_j}=\frac{1}{\sqrt{\e}b_j}\Big(-iA_{z_j}+\frac{1}{\sqrt{\t_j}}
a^\dagger_j\Big), \quad
\bar C_j = C_j^{\dagger}=\bar C_{\bar z_j}=\frac{1}{\sqrt{\e}b_j}\Big(
iA_{\bar z_j}+\frac{1}{\sqrt{\t_j}}a_j \Big).
\end{eqnarray}
In addition to the equations of motion, the gauge condition $A_0 =0$ induces
the Gauss law constraint
\begin{equation}
\label{Gauss}
\sum_{i=1}^{p/2} \left( \left[ C_i , \pa_t\bar C_i\right]
+ \left[ \bar C_i, \pa_t C_i\right] \right) =0.
\end{equation}

On D$0$-D$p$, we can construct exact solitonic solutions by applying
``solution generating technique''~\cite{HKL}. This is defined by the
following ``almost gauge transformation'':
\begin{eqnarray}
\label{hkl}
C_i\to S_k^{\dagger} C_i S_k+\sum_{l=1}^{k}\xi_l^i | {p_l}\ran\lan{p_l}|,
\end{eqnarray}
where $|{p_l}\ran$ is orthogonal and normalized states of the oscillators,
and $S_k$ is an almost unitary operator, which is usually called a partial
isometry and satisfies
\begin{eqnarray}
S_k S_k^\dagger=1,~~~S_k^\dagger S_k=1-P_k,
\end{eqnarray}
where $P_k=\sum_{l=1}^{k} |p_l\ran\lan p_l|$ is a projection operator whose
rank is $k$. A typical example of the partial isometry is a shift operator,
given, for example, in \cite{HKL}. It has been argued that the complex
parameters $\xi_l^i$ represent the positions of the $k$
solitons~\cite{AGMS,GN,HO,F4,H}. We are now going to present another evidence
for this interpretation by showing that the fluctuation spectra of two
D0-branes on D$p$-branes are proportional to $|\xi_1^i-\xi_2^i|^2$
which is the distance between two solitons.

The transformation (\ref{hkl}) leaves the equation of motion and the
Gauss law constraint satisfied~\cite{HKL}, and generates the following
nontrivial solution from the vacuum solution $A_i=0$:
\begin{eqnarray}
C^{(0)}_i = \frac{1}{\sqrt{\bar b_i}} S^\dagger_k a^\dagger_i S_k
+\sum_{l=1}^{k}\xi_l^i | p_l\ran\lan p_l |,
\label{sol2}
\end{eqnarray}
which corresponds to the D-brane system of $k$ D0-branes on a D$p$-brane.
The parameters $\xi_l^i$ are arbitrary and are called the moduli of the
solitons.

Let us investigate small fluctuations around the exact solution~(\ref{sol2})
represented by
\begin{eqnarray}
\hs{-5}C_i\!\!&=&\!\! C^{(0)}_i+\delta C_i \nn
   \!\!&=&\!\! C^{(0)}_i+P_k A_iP_k+P_kW_i(1-P_k)+(1-P_k)\bar T_iP_k+
S_k^\dagger D_iS_k.
\label{fluc}
\end{eqnarray}
The mass matrix of the fluctuations is obtained by substituting (\ref{fluc})
into the action~(\ref{action}). Just for simplicity we study $k=2$ case and
focus on the fluctuations $A_i$ which correspond to 0-0 strings. The classical
solution is
\begin{eqnarray}
C_i^{(0)}=\frac{1}{\sqrt{\bar b_i}}S_2^{\dagger} a_i^\dagger S_2
+\xi^i_1|{p_1}\ran\lan{p_1}|+\xi^i_2|{p_2}\ran\lan{p_2}|,
\end{eqnarray}
and the fluctuations around it are written as
\begin{eqnarray}
C_i=\left(
\begin{array}{cc}
B_i & W_i\\
\bar{T}_i & S^\dagger_2 (a_i^\dagger/\sqrt{\bar b_i} +D_i)S_2
\end{array}
\right), \quad
B_i=\left(
\begin{array}{cc}
A^i_{11}+\xi^i_1 & A^i_{12}\\
A_{21}^i & A_{22}^i + \xi^i_2,
\end{array}
\right)
\end{eqnarray}
where $A_{jk}^{i}$ are the fluctuations that we are interested in.
The mass terms for the fluctuations in the Lagrangian are found to be
\begin{eqnarray}
{\cal L} &=& \sum_{i,j=1}^{p/2}\left\{2\vert \xi_1^j
-\xi_2^j\vert^2\vert A_{12}^i\vert^2
+2\vert \xi_1^j-\xi_2^j\vert^2\vert A_{21}^i\vert^2\right.\nn
&&\left.~~~~~-(\xi_1^i-\xi_2^i)(\xi_1^j-\xi_2^j)\bar{A}_{12}^i\bar A_{21}^j
-(\bar \xi_1^i-\bar \xi_2^i)(\bar \xi_1^j-\bar \xi_2^j)A_{12}^iA_{21}^j
\right.\nn
&&\left.~~~~~-(\xi_1^i-\xi_2^i)(\bar \xi_1^j-\bar \xi_2^j)\bar{A}_{12}^i
A_{12}^j -(\xi_1^i-\xi_2^i)(\bar \xi_1^j-\bar \xi_2^j)A_{21}^j\bar A_{21}^i
\right\}.
\end{eqnarray}
Diagonalizing this mass matrix, we get the mass spectra in terms of the
properly normalized coordinates $x^i_l \equiv \sqrt{2\bar b\theta} \xi^i_l$:
\begin{eqnarray}
0,~~~\frac{\e}{(2\pi\a')^2}\sum_{i=1}^{p/2} |x^i_1-x^i_2|^2.
\end{eqnarray}
It can be shown that the zero eigenvalue corresponds to the unphysical
mode specified by the Gauss law~\p{Gauss}. The other eigenvalues show
that the open string stretched between two D0-branes has the mass
proportional to $|x_1-x_2|$. This is consistent with the picture that
the parameters $\xi_1^i$ and $\xi_2^i$ correspond to the positions of the two
D0-branes with strings stretched between them, and the string tension is
given by $\sqrt{\e}/2\pi\a'$, as we expected.

Though we have explicitly checked this interpretation for $k=2$, there
should be no difficulty in extending our method to arbitrary $k$.

\section{Low-energy scattering of D0-D0 on D$p$-branes}

As an interesting application of our results, let us discuss low-energy
scattering of two BPS D0-branes on D$p$-branes. Since the solutions we are
considering are BPS without static force, the scattering can be described by
the geodesic in the moduli space~\cite{M}. To be more explicit, let us consider
two D0-branes on D4-branes whose moduli are those of $U(1)$ two instantons.
To examine the scattering of D0-branes, it is necessary to know the metric
of the moduli for the relative positions of D0-branes. This can be read off
from the kinetic terms in the action of the D$p$-branes when the soliton
solutions with time-dependent positions are substituted~\cite{AH,GM}. It
turns out that the metric for the D0-branes generated by solution generating
technique is flat, so that the scattering is trivial. However, it is possible
to construct more general BPS solitons by using ADHM construction. The metric
of the BPS instanton moduli is then equivalent to the solution space of
ADHM equation. We can determine the metric of the moduli space from the
general solutions of ADHM equation. This enables us to derive the geodesic
on it and discuss the classical scattering of two D0-branes. We discuss
this problem for each D0-D$p$ system separately.

\subsection{D0-D4 System}

First let us consider the system of $k$ D0-branes on $N$ D4-branes
with background $B$-field. This system corresponds to self-dual
$G=U(N)~k$-instanton on noncommutative ${\bf R}^4$.
The moduli space of the system is described by the deformed ADHM equation:
\begin{eqnarray}
\label{adhm}
&&[\Phi_1,\Phi_1^\dagger]+[\Phi_2,\Phi_2^\dagger]+II^\dagger-J^\dagger J
 = \zeta,\nn
&&{[\Phi_1,\Phi_2]}+IJ=0.
\end{eqnarray}
where $\Phi_i~(i=1,2)$ and $I,J^\dagger$ are $k\times k$ and $k\times N$
matrices and correspond to 0-0 strings and 0-4 strings, respectively.
The real parameter $\zeta$ is given in terms of the noncommutativity
parameters as $\zeta=\theta_1-\theta_2$. Note that the self-dual case
corresponds to $\zeta=0$.

To determine the solution space of ADHM equation (\ref{adhm}), we have to
find its general solutions. Those for $G=U(1)$ and $k=2$ are found in
\cite{LTY} to be
\begin{eqnarray}
\Phi_i = w^{\rm c}_i +\frac{w^{\rm r}_i}{2}
\left(\begin{array}{cc} 1 & \displaystyle\sqrt{\frac{2b}{a}} \\ 0 & -1
\end{array} \right),~~~
I = \sqrt{\zeta}\left(\begin{array}{c}\sqrt{1-b} \\ \sqrt{1+b}\end{array}
\right),~~~J=0,
\end{eqnarray}
where
\begin{eqnarray}
a=\frac{\vert w^{\rm r}_1\vert^2
+\vert w^{\rm r}_2\vert^2}{2\zeta},~~~
b=\frac{1}{a+\sqrt{1+a^2}}.
\end{eqnarray}
The complex parameters $w^{\rm c}_i \sim (\xi_1^i+\xi_2^i)/2$
and $w^{\rm r}_i\sim \xi_1^i - \xi_2^i$ correspond to
the center of mass and relative positions, respectively.

The metric of the moduli space is also derived in \cite{LTY}
by considering infinitesimal gauge transformation $\delta$
and linearized Gauss law:
%\footnote{The linearized Gauss law given in \cite{LTY}
%can be derived from the ADHM equation:
%$$
%\mbox{Tr}(\nabla^\dagger\delta\nabla-\delta\nabla^\dagger\nabla)=0.
%$$}
\begin{eqnarray}
ds^2=2 {\rm tr}(\d \Phi_1\d \Phi_1^\dagger+\d \Phi_2 \d \Phi_2^\dagger).
\end{eqnarray}
The metric naturally decomposes into the parts of the center of mass and the
relative motions. The latter part turns out to be
\begin{eqnarray}
\label{rel}
ds^2_{\rm rel}
=f(r)\Big(dr^2+\frac14 r^2 \sigma_z^2\Big)
+\frac14 f(r)^{-1}r^2(\sigma_x^2+\sigma_y^2),
\end{eqnarray}
where
\begin{eqnarray}
f(r)=\sqrt{1+\frac{4\zeta^2}{r^4}},
\end{eqnarray}
and
\begin{eqnarray}
\sigma_x&=&-\sin \psi d \theta+\cos \psi \sin \theta d\varphi,\nn
\sigma_y&=&\cos \psi d \theta+\sin \psi \sin \theta d\varphi,\nn
\sigma_z&=& d \psi + \cos \theta d\varphi,
\end{eqnarray}
are the $SU(2)$ invariant one-forms. We note that the metric (\ref{rel})
becomes flat in the case $\zeta=0$, that is, if noncommutativity parameter
$\t_i$ is self-dual. This is the case for the BPS solitons generated by
the solution generating technique, and hence we again find here that the
scattering is trivial in that case.

Now let us find the geodesic on the moduli space. The geodesic equation
is given as the equation of motion following from the action:
\begin{eqnarray}
I=m\int d\tau~g_{\mu\nu}^{\rm rel}\frac{du^\mu}{d\tau}\frac{du^\nu}
{d\tau},
\end{eqnarray}
where the metric $g_{\mu\nu}^{\rm rel}$ is read from (\ref{rel})
and $u^\mu=(r,\theta,\psi,\varphi)$ are the coordinates of the moduli space.
The variational equation $\delta I=0$ yields
\begin{eqnarray}
\frac{1}{f(r)}\dot r^2+\frac{r^2}{4f(r)}(\dot \psi + \cos \theta \dot\varphi)^2
+\frac14 f(r)r^2(\dot \t^2+\sin^2 \theta \dot\varphi^2)&=&E, \nn
\frac{r^2}{f(r)}(\dot\psi + \cos \theta \dot\varphi)&=& L, \nn
L\cos \theta+f(r)r^2\sin^2\theta\dot\varphi^2&=&C, \nn
\frac{d}{d\tau}(f(r)r^2\dot\t) +L\sin \theta \dot\varphi
-f(r)r^2 \sin\theta\cos\theta\dot\varphi^2&=&0,
\end{eqnarray}
where the dot stands for the differentiation with respect to the parameter
$\tau$ describing scattering process (which can be regarded as time), and
$E,L$ and $C$ are the integration constants.
The solution of our interest for these equations is
\begin{eqnarray}
\dot\varphi=0,~~~\dot\t=0,~~~\dot\psi =L \frac{f(r)}{r^2},~~~\dot r ^2+V(r)=0,
\end{eqnarray}
where
\begin{eqnarray}
V(r)=f(r)\left(\frac{L^2 f(r)}{4r^2}-E\right).
\end{eqnarray}
Our problem thus reduces to the classical dynamics for the scattering of
zero-energy particles with potential $V(r)$. Introducing the impact parameter
$\rho=L/2\sqrt{E}$ and the turning point $r=r_0$ defined by $V(r_0)=0$,
we get
\begin{eqnarray}
\frac{dr}{d\psi}=\frac{r}{2}\sqrt{\frac{r^2}{\rho^2f(r)}-1}.
\end{eqnarray}
Therefore the exit angle is derived as
\begin{eqnarray}
\frac{\psi_{\rm exit}}{2}
=\int_{y_0}^{\infty}
\frac{dy}{y\displaystyle\sqrt{\frac{2\zeta y^2}{\rho^2\sqrt{y^2+1}}-1}},
\end{eqnarray}
where $y=r^2/2\zeta$ and $y_0=r_0^2/2\zeta$. The angle $\psi$ covers the
whole space twice for the range $0\leq \psi \leq 2\pi$, so it is more
convenient to call $\psi_{\rm exit}/2$ the exit angle. It is plotted
as a function of the logarithm of the impact parameter in Figure 1.
%\begin{figure}[htbp]
%\begin{center}
%\includegraphics[width=7.5cm]{scatter.eps}
%\end{center}
%\label{scatter}
%\caption{exit angle versus log of impact parameter}
%\end{figure}
\begin{figure}[htbp]
\begin{center}
\includegraphics[width=7.5cm]{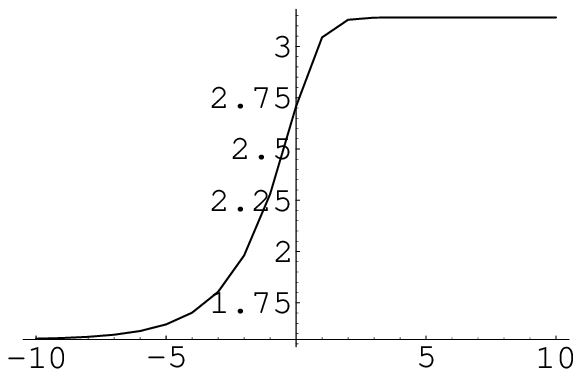}
\put(-153,115){{\footnotesize $\psi_{\rm exit}/2$}}
\put(-56,19){{\footnotesize $\log\rho$}}
\end{center}
\label{scatter}
\caption{exit angle versus log of impact parameter}
\end{figure}

{}From this figure, we find that the exit angle is $\pi$ for large impact
parameter and gradually decreases if the impact parameter is decreased.
In particular, the exit angle for the head-on collision is $\pi/2$,
as is the case for monopoles and GMS solitons. We again note that for
$\zeta=0$, the scattering is trivial, which means that the D0-branes
generated by solution generating technique scatter trivially.
Our result indicates that the more general background $B$-field makes the
scattering nontrivial.

\subsection{D0-D6 System}

Next we consider the system of $k$ D0-branes on $N$ D6-branes
with background $B$-field.
The moduli space of the system is determined in \cite{W} by
\begin{eqnarray}
\label{6adhm}
&&[\Phi_1,\Phi_1^\dagger]
+ [\Phi_2,\Phi_2^\dagger]
+ [\Phi_3,\Phi_3^\dagger]+I I^\dag = \zeta, \nn
&&[\Phi_1,\Phi_2] = 0,~~~[\Phi_2,\Phi_3] = 0,~~~[\Phi_3,\Phi_1] = 0,
\end{eqnarray}
where $\Phi_i~(i=1,2,3)$ and $I$ are $k\times k$ and $k\times N$ matrices
and correspond to 0-0 and 0-6 strings, respectively, as in D0-D4 system.
The real parameter $\zeta$ is a FI parameter and depends on the background
$B$-field. Only when $\zeta \geq 0$, eq.~(\ref{6adhm}) has solutions.

Since the general solution of (\ref{6adhm}) has not been found,
let us investigate special solutions for $G=U(1)$ case. If we restrict
$\Phi_3 = w_3^{{\rm c}}$, then eq.~(\ref{6adhm}) reduces to ADHM
equation~(\ref{adhm}), where $w_i^{{\rm c}}$ represents the center of mass
coordinate. Hence the moduli space of this simple solution is the same as
that of D0-D4 system and the scattering process will be the same as D0-D4
case,\footnote{In D0-D0 scattering on a D4 system, we have restricted it
to the $r$-$\psi$ plane by taking $\dot{\theta}=\dot{\varphi}=0$,
which might justify the discussion here.}
implying that the exit angle for the head-on collision is generally
$\pi/2$ and only $\zeta\neq 0$ leads to nontrivial scattering.
If $\zeta=0$, in fact, the general solution is found as
$\Phi_i=\mbox{diag}_l(\xi^i_l),~I=0$ and
the metric of the moduli becomes flat.

\subsection{D0-D8 System}

Finally consider $k$ D0-branes on $N$ D8-branes with background $B$-field.
The equation for the moduli space is again not known explicitly. However
there exists an eight-dimensional ADHM construction which gives rise to
some class of eight-dimensional instantons~\cite{CGK}. We examine the
eight-dimensional ADHM equations on noncommutative ${\bf R}^8$~\cite{OK}
and focus on the subspace of the moduli space and the corresponding
scattering process.

The eight-dimensional ADHM equations are given by~\cite{CGK,OK}
\begin{eqnarray}
\label{8adhm}
&&[\Phi_1,\Phi_1^\dagger] + [\Phi_2,\Phi_2^\dagger] + I I^\dagger
- J^\dag J = \zeta, \nn
&&[\Phi_1,\Phi_2] + IJ = 0, \nn
&&[\Phi_1,\Phi_3^\dagger] + [\Phi_2,\Phi_4^\dagger]
+ I K^\dagger - L^\dag J = 0, \nn
&&[\Phi_1,\Phi_4] + [\Phi_3,\Phi_2] + I L +K J = 0, \nn
&&[\Phi_3,\Phi_3^\dagger] + [\Phi_4,\Phi_4^\dagger]
+ K K^\dagger - L^\dag L = 0, \nn
&&[\Phi_3,\Phi_4] + K L = 0,
\end{eqnarray}
where $\Phi_i~(i=1,2,3,4)$ and $I,J,K,L$ are $k\times k$ and $k\times N$
matrices, respectively. The parameter $\zeta$ depends not only on the
background $B$-field but also on the matrices $\Phi_3,\Phi_4,K,L$.
These equations are the (restricted) D-flatness conditions in the worldvolume
theory on the D0-branes, and then $\Phi_i~(i=1,2,3,4)$ and $I,J,K,L$
correspond to 0-0 and 0-8 strings, respectively.

As in the case of D0-D6, it is difficult to solve these equations fully.
Hence we look for special solutions. A solution is obtained by putting
$\Phi_3 = w_3^{{\rm c}},~ \Phi_4 = w_4^{{\rm c}},~K=L=0$. Then
eq.~(\ref{8adhm}) reduces to ADHM equation~(\ref{adhm}), and the problem
is similar to the D0-D4 systems. By the same reasoning as D0-D6 system,
we conclude that the scattering of D0-D0 on D8-branes would be the same
as that of D0-D4 and the scattering would occur at right angle for the
head-on collision.

\section{Conclusions and Discussions}

We have discussed moduli space of D0-D$p$-brane systems. We have shown
that the moduli parameters in solution generating technique represent the
positions of the solitons by examining the fluctuation spectra corresponding
to open strings between D0-branes. As an interesting application of our
results, we have also examined the scattering process of D0-branes in
the D$p$ effective theory for arbitrary noncommutativity parameters without
approximation. The exit angle is determined as a function of the impact
parameter, and in particular it turns out to be $\pi/2$ for the head-on
collision, which is a universal result in low-energy soliton scattering.
If $\zeta=0$ which corresponds to self-dual solutions and those constructed
by solution generating technique, the scattering becomes trivial.
Hence the existence of the general background $B$-field is important to
render the scattering nontrivial.

We have some comments on the universal results of such two soliton
scatterings. In all cases, the two solitons are treated as bosons and
the moduli spaces have ${\bf Z}_2$ symmetry. The metric (\ref{rel}) for
the D0-D4 system is in fact equivalent to Eguchi-Hanson metric~\cite{EH}
which is a resolution of the orbifold ${\bf C}^2/{\bf Z}_2$.
Similarly the moduli spaces of D0-D6 and D0-D8 systems for $G=U(1)$ are
considered to be resolutions of the orbifolds ${\bf C}^3/{\bf Z}_2$
and ${\bf C}^4/{\bf Z}_2$, respectively. The boundary of Eguchi-Hanson
space at the infinite distance between two D0-branes is $S^3/{\bf Z}_2$.
$S^3$ has a Hopf-fibration whose fiber is $S^1$ with the coordinate $\psi$.
The ${\bf Z}_2$ symmetry would give rise to the right angle scattering for
the head-on collision. Similarly the boundary of the moduli for D0-D8 system
is $S^7/{\bf Z}_2$. $S^7$ also has a Hopf-fibration whose fiber is $S^3$
and this part corresponds to the boundary of Eguchi-Hanson space.
This is why the moduli space of D0-D8 system contains that of D0-D4 system
as is seen in subsection 3.3, and the universal scattering behavior is
expected because of the ${\bf Z}_2$ symmetry.

There is another important BPS D-brane system corresponding to BPS monopoles:
$k$ D1-branes ending on $N$ D3-branes~\cite{Diac}. For $N=1$ and 2, the
moduli space is unchanged by the presence of $B$-field on the
D3-branes~\cite{GN_m}. Hence the scattering process is all the same as
commutative case; especially the noncommutative $U(2)$ monopoles scatter at
right angle for the head-on collision, and get converted into noncommutative
dyons.

Note added. It was pointed out to us that scattering of noncommutative
solitons was also discussed in \cite{LP}.

%\newpage
%\vskip3mm
\noindent
{\bf Acknowledgments}
\vskip3mm
We would like to thank K. Hashimoto and R. Gopakumar for valuable discussions,
which motivated the present work. We also acknowledge the Summer Institute
2001 at Fuji-Yoshida, where this work was begun. Thanks are also due to
Y.~Imamura, H.~Kanno, K.~Ohta, T.~Takayanagi and T.~Uesugi for discussions and
K.~Ichikawa and T. Shindou for help with computer.
MH thanks the YITP at Kyoto University for the hospitality during his stay.
The work of MH was supported in part by the Japan Securities
Scholarship Foundation (\#12-3-0403). That of NO was supported in part
by a Grant-in-Aid for Scientific Research No. 12640270, and by a
Grant-in-Aid on the Priority Area: Supersymmetry and Unified Theory of
Elementary Particles.

\end{document}